\documentstyle[prb,aps,epsfig,floats]{revtex}

\begin{document}
\title{Critical temperature oscillations in magnetically coupled superconducting mesoscopic loops}
\author{M. Morelle$^1$, V. Bruyndoncx$^1$, R. Jonckheere$^2$ and V.V. Moshchalkov$^1$}
\address{$^1$Laboratorium voor Vaste-Stoffysica en Magnetisme, K.U. Leuven,
Celestijnenlaan 200D, B-3001 Leuven, Belgium \\
$^2$Interuniversity
Micro-Electronics Center, Kapeldreef 75, B-3001 Leuven, Belgium}

\date{\today}
\maketitle
\begin{abstract}
We study the magnetic interaction between two superconducting
concentric mesoscopic Al loops, close to the
superconducting/normal phase transition. The phase boundary is
measured resistively for the two-loop structure as well as for a
reference single loop. In both systems Little-Parks oscillations,
periodic in field are observed in the critical temperature $T_c$
versus applied magnetic field $H$. In the Fourier spectrum of the
$T_c(H)$ oscillations, a weak 'low frequency' response shows up,
which can be attributed to the inner loop supercurrent magnetic
coupling to the flux of the outer loop. The amplitude of this
effect can be tuned by varying the applied transport current.

\end{abstract}

\section{Introduction}
In 1962, Little and Parks\cite{LittleParks} measured a mesoscopic
superconducting cylinder in an axial magnetic field. The
superconducting critical temperature $T_c(\Phi/\Phi_0)$ showed
oscillations periodic in the normalized flux, with the period
corresponding to the superconducting flux quantum $\Phi _0=h/2e$.
These oscillations in $T_c(\Phi/\Phi_0)$ are a straightforward
consequence of the fluxoid quantization constraint, which was
introduced by F. London\cite{London}. Fluxoid quantization can be
easily understood by integrating the second Ginzburg-Landau (GL)
equation for the supercurrent\cite{Tinkhambook,deGennes}
\begin{equation}
\vec{j}=\frac{2e}{m^\star}\left|\Psi\right|^2\left(\hbar
\vec{ \nabla} \delta -2 e \vec{A}\right)=2 e \left|\Psi\right|^2
\vec{v}
\label{Eq:Supercurrent}
\end{equation}
along a closed contour. Here, $\vec{j}$ is the supercurrent
density, $\vec{v}$ is the superfluid velocity, $\delta$ is the
phase of the complex order parameter $\Psi=\left|\Psi\right|e^{i
\delta}$, and $\vec{A}$ is the magnetic vector potential.
Integration along an arbitrary closed contour yields the following
equation:
\begin{equation}
\Phi ' \equiv \Phi + \frac{1}{2e}\oint m^\star \vec{v}\cdot
d\vec{l}=N\Phi_0, \label{Eq:Flux}
\end{equation}
where the fluxoid $\Phi '$ is quantized in units of $\Phi_0=h/2e$
and $\Phi$ is the applied flux threading the area inside the
contour. The integer number $N$ is the phase winding number, or
also called the fluxoid quantum number, counting the number of
flux quanta $\Phi_0$ penetrating the enclosed area. When the
applied flux $\Phi$ is not equal to integer times the flux quantum
$\Phi_0$, a supercurrent $j$ has to be generated in order to
fulfill Eq. (\ref{Eq:Flux}).

For a superconducting ring of radius $r$, made of wires of
vanishing width ($w=0$), a one-dimensional (1D) GL model can be
used to describe the onset of superconductivity. The relative
$T_c$ variations can be written as:
\begin{equation}
\frac{T_{c0}-T_c(\Phi/\Phi_0)}{T_{c0}}=\frac{\xi^2(0)}{r^2}\left(N-\frac{\Phi}{\Phi_0}\right)^2,
\label{Eq:Wire}
\end{equation}
where $\xi(0)$ is the coherence length at zero temperature,
$T_{c0}$ is the critical temperature in zero field , and the
integer number $N$ is chosen to maximize the critical temperature
$T_c(\Phi)$. For each fluxoid quantum number $N$, the critical
temperature $T_c(\Phi/\Phi_0)$ has a parabolic shape. The
Little-Parks (LP) oscillations appear due to the transitions from
the integer value $N$ to $N+1$, at a half integer value of
$\Phi/\Phi_0$. For $\Phi/\Phi_0=N$, no supercurrent flows in the
ring, while for $\Phi/\Phi_0=N+1/2$, the supercurrent reaches a
maximal value and changes sign. The maximum normalized variation
of the critical temperature $\Delta T_c(\Phi/\Phi_0)/T_{c0}$ is
$\xi^2(0)/4r^2$, as can be evaluated from Eq. (\ref{Eq:Wire}). To
observe these quantization effects experimentally, structures with
the radius of the loop of the order of the coherence length
$\xi(0)$ should be used.

Since the pioneering work of Little and Parks, single mesoscopic
superconducting loops and cylinders has been largely studied.
Recently, Zhu $et$ $al.$\cite{Zhu} studied the flux state in two
magnetically coupled mesoscopic normal loops. The magnetic
coupling of an array of normal\cite{Wang}  and
superconducting\cite{Davidovic} loops has also been studied. In
those cases the loops are electrically isolated from each other
and can only interact magnetically.

Our work will focus on the magnetic coupling of two concentric
superconducting loops. The modification of the phase boundary
$T_c(H)$ of the outer loop due to magnetic coupling with the inner
loop will be studied. Magnetic coupling  effects between two loops
are potentially very important since for the two loops made from
different materials, a new unusual effect of enhancing $T_{c1}(H)$
due to a higher $T_{c2}(H)$ can be expected.

The GL free energy for such system can be written as follows:
\begin{eqnarray}
{\cal F}_s = {\cal
F}_n+V_i\left(\alpha\left|\Psi_i\right|^2+\beta\left|\Psi_i\right|^4+\frac{m^\star
v_i^2}{2}\left|\Psi_i\right|^2\right)\nonumber\\
+V_o\left(\alpha\left|\Psi_o\right|^2+\beta\left|\Psi_o\right|^4+\frac{m^\star
v_o^2}{2}\left|\Psi_o\right|^2\right)\nonumber\\
+L_iI_i^2+L_oI_o^2+MI_iI_o,
\label{Eq:Freeenergy}
\end{eqnarray}
with ${\cal F}_n$ the total free energy in the normal state,
$\alpha$ and $\beta$ the expansion coefficients, $m^\star$ the
mass of a Cooper pair and $L$ and $M$ the self- and mutual
inductance. $V$ is the volume of the loop, $\Psi$ is the order
parameter, $v$ is the velocity of the superfluid and $I$ is the
supercurrent. The indexes $i$ and $o$ refer to the inner and outer
loop, respectively. The superfluid velocities of the two loops are
determined from the fluxoid quantization constraint:
\begin{equation}
\begin{array}{ll}
\displaystyle v_i=\frac{\hbar}{m^\star
r_i}\left(N_i-\frac{\Phi+I_oM}{\Phi_0}\right)
\\
\\
\displaystyle v_o=\frac{\hbar}{m^\star
r_o}\left(N_o-\frac{\Phi+I_iM}{\Phi_0}\right).
\end{array}
\label{Eq:Velocity}
\end{equation}
To solve this equation, the free energy must be minimized with
respect to variations in $\Psi_i$, $\Psi_o$, $I_i$ and $I_o$. The
free energy ${\cal F}_s$ contains in this case the term
$MI_iI_o\propto M\left|\Psi_i\right|^2 \left|\Psi_o\right|^2$
responsible for the mixing of the two individual order parameters
$\Psi_i$ and $\Psi_o$.

\section{Experimental results and
discussion}
\begin{figure}[ht]
\centering
\includegraphics*[width=8cm,clip=]{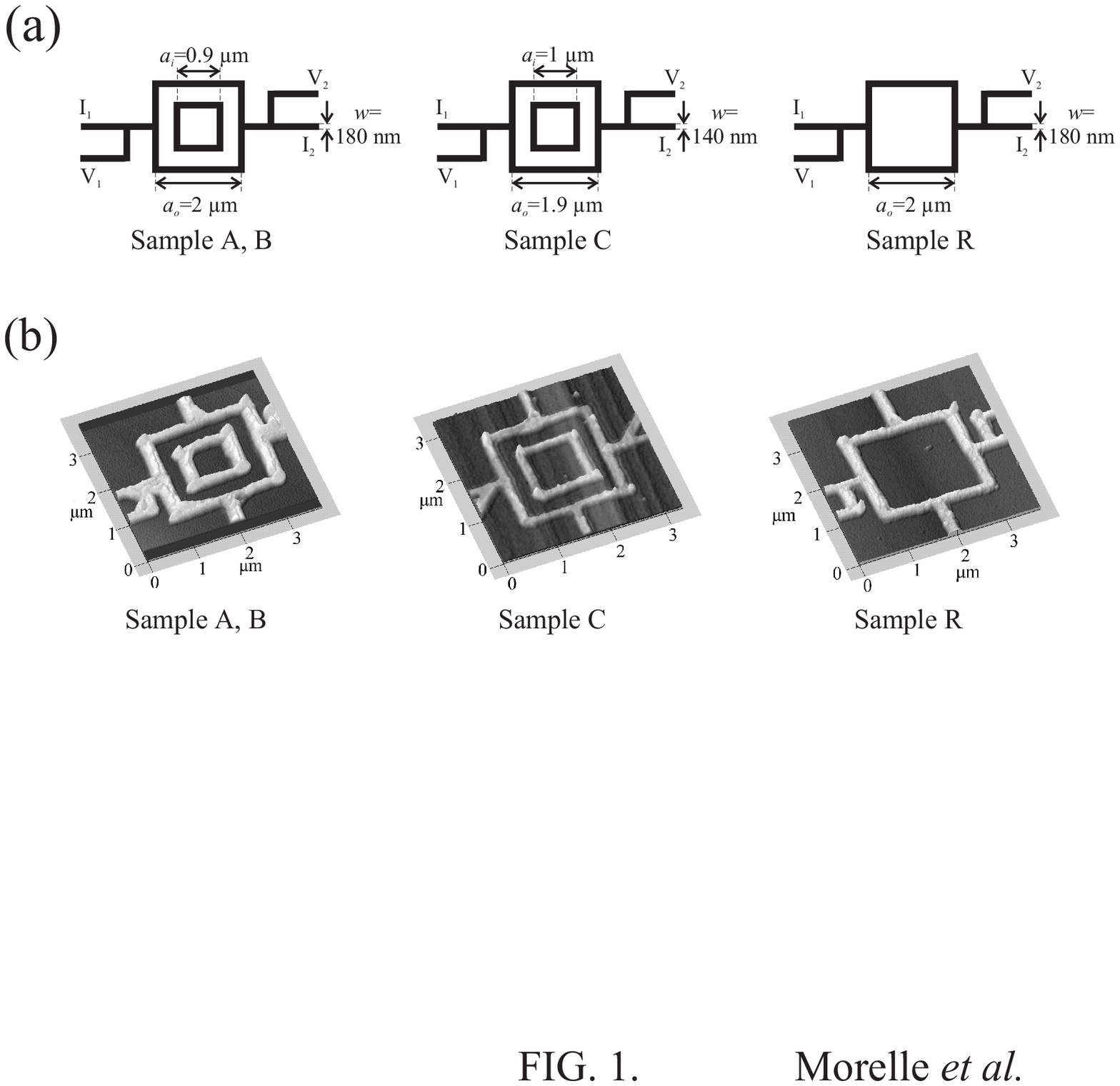}
\caption{(a) A schematic drawing of samples A, B and C, and sample
R. I$_1$ and I$_2$ are the contacts for the transport current;
V$_1$ and V$_2$ are the voltage probes. $a_i$ and $a_o$ are the
width of the inner, resp. outer loop, measured from middle to
middle, and $w$ is the wire width of the loop and of the
connecting wires. (b) AFM image of the double loop (sample A, B
and C) and of the single loop (sample R).} \label{Fig:1}
\end{figure}

We present the results of transport measurements carried out on
two different types of mesoscopic Al structures (Fig.
\ref{Fig:1}). The first type of sample is composed of two
concentric loops, with the outer loop being electrically connected
to the experimental set-up in order to perform four-points
resistance measurements. Three samples of this type have been
studied: sample A and sample B with the same thickness and the
same dimensions of the loops and sample C with a smaller thickness
and slightly different dimensions of the loops. The reference
sample (sample R) is analogous to the first structure, but without
the inner loop. All the samples discussed, except sample C, are
evaporated in the same run. All samples have been prepared by
thermal evaporation of 99.9999\% pure Al on a SiO$_2$ substrate.
The patterns are defined using electron beam lithography on a
bilayer of PMMA/PMMA-coPMMA resist before the deposition of an
aluminum film with a thickness $\tau$=50 nm and $\tau$=28 nm for
sample A, B and R and for sample C, respectively. After the
evaporation, the lift-off was performed using dichloromethane. In
Fig. \ref{Fig:1}a, the geometry and the dimensions of the
different structures are shown. The thickness and the lateral
dimensions of the samples have been characterized by the X-ray
diffraction on a co-evaporated plane film and AFM (see Fig.
\ref{Fig:1}b), respectively. The wire width of the loop and of the
connecting wires has been determined from SEM investigations. The
superconducting/normal phase boundaries are obtained from
transport measurements, carried out with a transport current $I_t$
flowing through the outer loop. The phase boundary is measured
holding the resistance at a fixed resistive criterion (we used the
criterion $R_n/2$, with $R_n$ the resistance in the normal state).
This is achieved using an electronic feedback circuit. Once a
required temperature stability is obtained, the magnetic field is
swept at a very slow rate (with a typical frequency of 20
$\mu$Hz). To improve the signal-to-noise ratio a PAR 124A lock-in
amplifier has been used, operating at 27.7 Hz.

The width of the wires of the studied superconducting loops
determines the parabolic background of the $T_c(H)$ phase
boundary\cite{Tinkham}:
\begin{equation}
\frac{T_{c0}-T_c(H)}{T_{c0}}=\frac{\pi^2}{3}\left(\frac{w\xi(0)\mu_0H}{\phi_0}\right)^2,
\label{Eq:PhaseB}
\end{equation}
where $w$ is the width of the wires. Eq. (\ref{Eq:PhaseB}) also
describes the $T_c(H)$ line for a superconducting thin film of
thickness $w$ subjected to a magnetic field parallel to the film
plane. In a practical situation, the phase boundary of a
superconducting loop of finite wire width will show a parabolic
background (see Eq. (\ref{Eq:PhaseB})). The suppression of $T_c$
can be written as the sum of two components: an oscillatory term
as described by Eq. (\ref{Eq:Wire}), and a monotonic term (Eq.
(\ref{Eq:PhaseB})). The coherence length $\xi(0)$ of the samples
can be determined from the parabolic background of the phase
boundary, since the width of the wires $w$ is a known parameter. A
second method to evaluate the coherence length is to determine
$\xi(0)$ from the slope of $T_c(H)$ of a co-evaporated macroscopic
reference film. A third method is based on the dirty limit of the
GL theory\cite{Tinkhambook}, where from the known value of the
elastic mean free path $l$, the coherence length
$\xi(0)=0.86\sqrt{\xi_0l}$ is obtained, using the clean limit
value $\xi_0$=1.6 $\mu m$ for Al. The results of the three methods
are summarized in table \ref{table:param}. The difference between
the $\xi(0)$ values calculated with the three different methods
can be partially explained by the rather broad error margins on
the wire width $w$.
\begin{figure}[ht]
\centering
\includegraphics*[width=8cm,clip=]{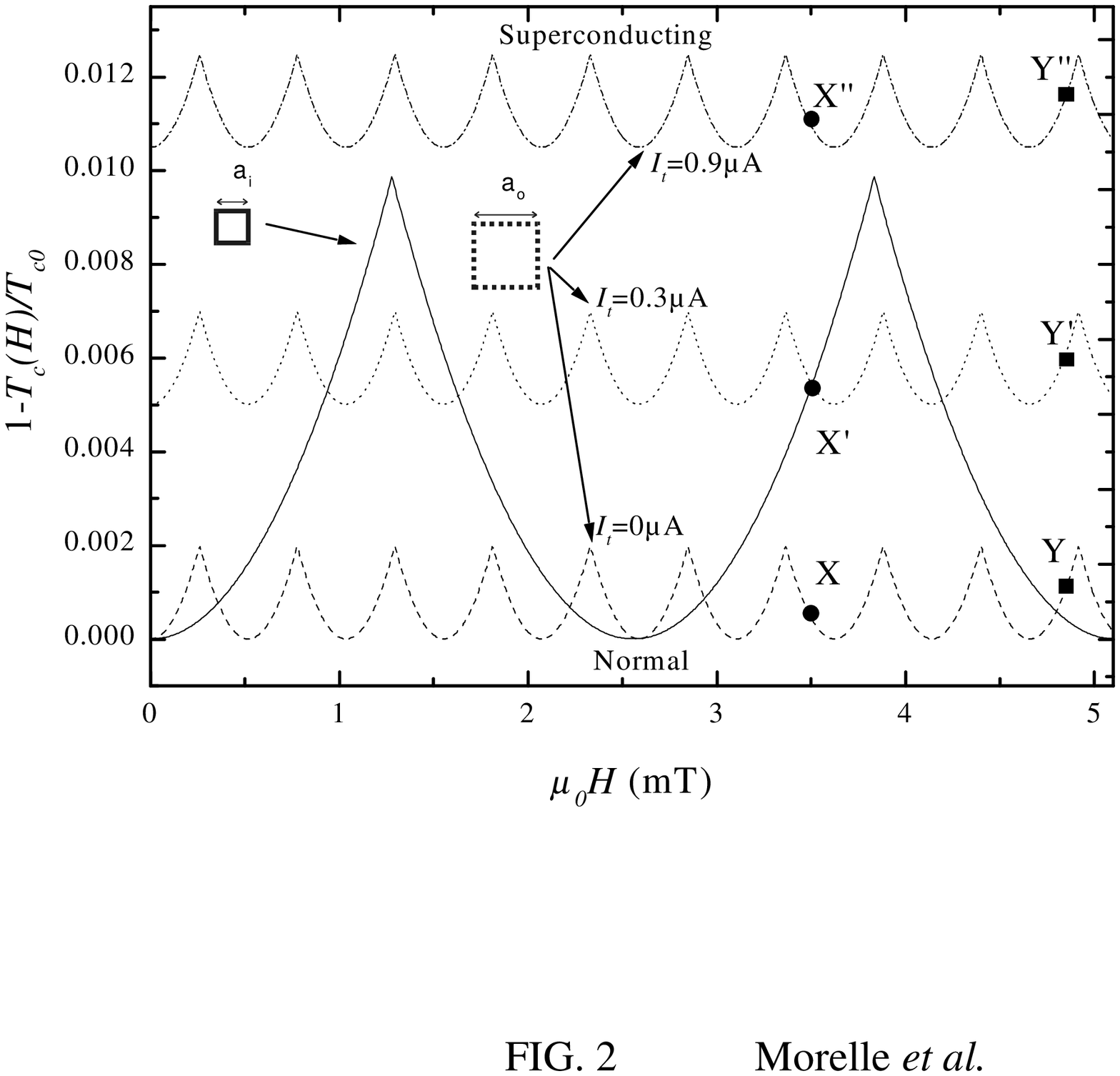}

\caption{Calculated phase boundary $T_c(H)$ of the inner (solid
line) and outer loop (dashed, dotted, and dashed-dotted line) for
two periods of the inner loop, without magnetic interaction, using
the dimensions of sample A and B. The phase boundary of the outer
loop is shown for three different transport currents $I_t$:
$I_t$=0 $\mu$A: dashed line, $I_t$=0.3 $\mu$A dotted line, and
$I_t$ =0.9 $\mu$A dashed-dotted line. The coherence length
$\xi(0)$=103 nm is taken. The shift of the $T_c(H)$ curves with
increasing $I_t$ are estimated from ref\cite{Strunk,Romijn}:
$I_c=I_{c0}(T_{c0}-T_c)^{3/2}$,with $I_{c0}\approx$ 550 $\mu$A.}
 \label{Fig:2}
\end{figure}

\begin{table*}[!t]
\begin{tabular}{ccccc}
sample & R & A & B & C \\   \tableline

$\tau$ (nm) & 50 & 50 & 50 & 28\\

$w$ (nm) & 180 & 180 & 180 & 140 \\

$a_o$ ($\mu$m) & 2 & 2 & 2 & 1.9 \\

$a_i$ ($\mu$m)  & --& 0.9 & 0.9 & 1 \\

$T_{c0}$ (K)  &1.324 & 1.326 & 1.327 & 1.363 \\

R$_\Box$ ($\Omega$)=$\rho/\tau$ (4.2 K) &0.65 &0.72 &0.76 & 1.36
\\

$l_{el}$ (nm) &12.3 &11 &10.6 & 10.5 \\

$\xi$(0) (Eq. (\ref{Eq:PhaseB})) (nm)& 103 & 105 & 102
 & 117 \\

$\xi$(0) Ref. Sample (nm)& 128 & 128 & 128 & -- \\

$\xi$(0) dirty limit (nm)& 114 & 112 & 120 & 112 \\

$L_i$ (pH) & -- & 1.6 & 1.6 & 1.9 \\

$L_o$ (pH) & 4.5 & 4.5 & 4.5 & 4.6 \\

$M$ (pH) & -- & 22 & 22 & 21\\

\end{tabular}
\caption{Material parameters for the measured samples}
\label{table:param}
\end{table*}

In Fig. \ref{Fig:2}, the theoretical phase boundaries are shown
for the two single loops when they are not coupled magnetically:
one using the dimensions of the inner loop to calculate the phase
boundary (solid line), and the other using the size of the outer
loop (lowest dashed line), thus corresponding to the reference
sample; this has been calculated from Eq. (\ref{Eq:Wire}) using
the dimensions of the loops summarized in table \ref{table:param}.
It should be mentioned that an increase along the vertical axis
corresponds to a decreasing temperature. The amplitude of the LP
oscillations is the larger the smaller the loop size (see Eq.
(\ref{Eq:Wire})). The period of the $T_c(H)$ oscillations for the
inner and outer loop is given by $\mu_0\Delta H_i=\Phi_0/S_i$, and
$\mu_0\Delta H_o=\Phi_0/S_o$, respectively with $S_i=a_i^2$ and
$S_o=a_o^2$ the enclosed areas (see Fig. \ref{Fig:1}). Points X
and Y in Fig. \ref{Fig:2} are situated at two fixed applied
magnetic field values on the $T-H$ phase boundary for the outer
loop. In the case of point X, the inner loop is in the normal
state, thus the inner loop carries no supercurrent. In this
situation, no flux is coupled to the outer loop. For point Y, on
the contrary, the inner loop is in the superconducting state, and
a supercurrent will flow in both loops in order to satisfy the
respective fluxoid quantization constrains (Eq. (\ref{Eq:Flux})).
Under these conditions, due to the mutual inductance between the
two current loops (see Eqs. (\ref{Eq:Freeenergy}) and
(\ref{Eq:Velocity})), an influence of the fluxoid quantization in
the inner loop on the measured $T_c(H)$ phase boundary of the
outer loop is expected.

To extend the flux interval for which the inner loop remains
superconducting, a higher transport current $I_t$ can be applied
to the outer loop. The phase boundary $T_c(\Phi)$ of the outer
loop is schematically presented in Fig. \ref{Fig:2} by the dotted
($I_t$=0.3 $\mu$A) and the dashed-dotted ($I_t$=0.9 $\mu$A) lines.
We now follow through the fixed magnetic fields lines following
the points X$\rightarrow$X'' and Y$\rightarrow$Y'' by increasing
the transport current $I_t$. The point on the phase boundary of
the outer loop, with the same magnetic field value as X for
$I_t$=0, will cross the phase boundary of the inner loop in point
X' by increasing $I_t$. For a highest transport current, the inner
loop will be superconducting for each point on the phase boundary
of the outer loop (dashed-dotted line). The inner loop will be
deeper in the superconducting state, following the shift from Y to
Y'' while increasing the transport current. As a result, an
increase of $I_t$ will not only broaden the interval in which the
inner loop is superconducting, but also increase the supercurrent
in the inner loop.

Zhang $et$ $al.$\cite{Zhang} have calculated the self-flux for a
typical mesoscopic ring. From these calculations, we can find the
self-flux and the additional flux in the outer loop due to the
presence of the inner loop. The self-flux for the inner and outer
loop will be not larger than  0.4\% of the applied flux for
$T/T_c>0.99$. The additional flux $MI_i$ in the outer loop will be
less than 4\% in this temperature interval but can be higher than
20\% for $T/T_c<0.95$. All the measurement presented in this paper
are in the region $T/T_c>0.99$. A possibility to further increase
the supercurrent in the inner loop $I_i$ and thus the magnetic
coupling between the two loops would be the use of a material with
a higher $T_c$ for the inner loop.

\begin{figure}[htb]
\centering
\includegraphics*[width=8cm,clip=]{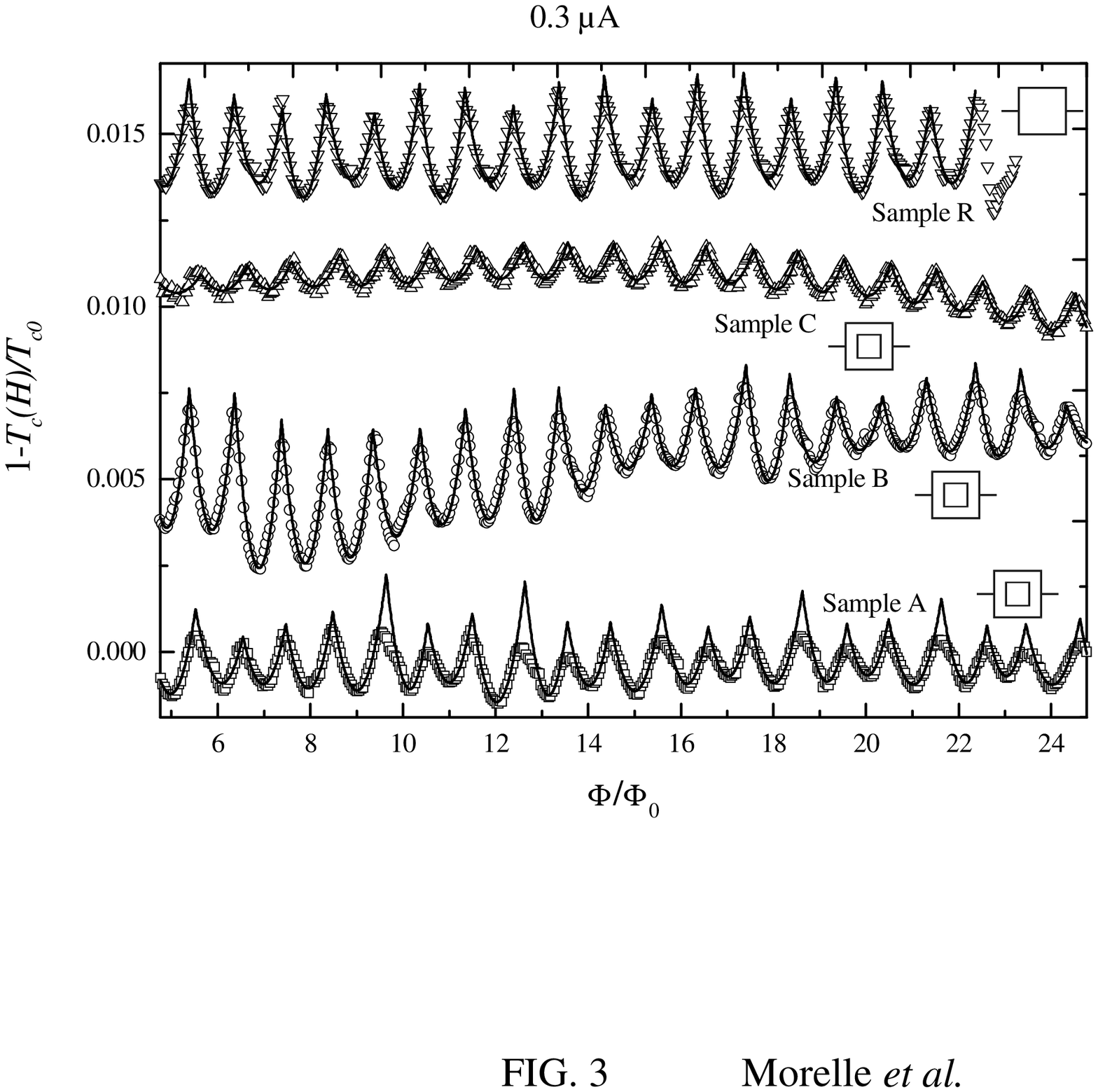}
\caption{Measured phase boundary $T_c(\Phi)$ after subtraction of
the parabolic background caused by the finite wire width of the
loops. The phase boundary is plotted in normalized units of the
flux $\Phi=S_o\mu_0 H$ threading the outer loop.  Within each
oscillation period, a parabolic function is fitted through the
data points in between the transition to a different fluxoid
quantum number $N_o$. The phase boundary of sample A ($\Box$),
sample B ($\bigcirc$), sample C ($\bigtriangleup$) and sample R
($\bigtriangledown$) is compared for the same transport current
$I_t$= 0.3$\mu$A.} \label{Fig:3}
\end{figure}

In Fig. \ref{Fig:3}, the phase boundaries of the samples A, B, C
and R ($\Box$: sample A, $\bigcirc$: sample B, $\bigtriangleup$:
sample C, $\bigtriangledown$: sample R) are shown for a ac
transport current $I_t$=0.3 $\mu$A rms after subtraction of the
monotonic parabolic background due to the finite width of the
strips (Eq. (6)). Further on, the flux $\Phi=S_o\mu_0H$ will refer
to the flux threading the surface of the outer loop. The
$T_c(\Phi)$ part below 5 $\Phi_0$ is not shown because the
experimental data were rather noisy in the low field region for
some measurements. The curves corresponding to different samples
are arbitrarily shifted in Fig. \ref{Fig:3}. The variation of the
critical temperature with increasing transport current can be
estimated from ref\cite{Strunk,Romijn} for zero field:
$I_c=I_{c0}(T_{c0}-T_c)^{3/2}$, with $I_{c0}\approx$550 $\mu$A and
$I_{c0}\approx$240 $\mu$A for sample A, B and R and for sample C,
respectively. The critical temperatures of the different samples
after extrapolation of the experimental results to $I_t$=0 are
given in table \ref{table:param}. For $I_t$=0.3 $\mu$A, the shift
of the zero field critical temperature $\Delta T_{c0}$ (used in
Fig. \ref{Fig:2}) is 7 mK (12 mK for sample C), for 0.5 $\mu$A: 9
mK and for 0.7 $\mu$A: 12 mK. Within each oscillation period, a
parabolic function is fitted through the data points in between
the transition to a different fluxoid quantum number
$N_o\rightarrow N_o+1$ (Eq. (\ref{Eq:Wire})). The value for the
self- and mutual inductance\cite{Duffin} for the different samples
are summarized in table \ref{table:param}.

In Fig. \ref{Fig:3}, a smaller amplitude of the oscillations is
observed for sample C, compared to the other samples. This smaller
amplitude is to be expected because of a smaller thickness of the
film, what would lead to a reduced mean free path and a smaller
coherence length value for sample C (see Eq. (\ref{Eq:Wire})). But
the coherence lengths are comparable for all four different
samples (see table \ref{table:param}).

\begin{figure}[htb]
\centering
\includegraphics*[width=8cm,clip=]{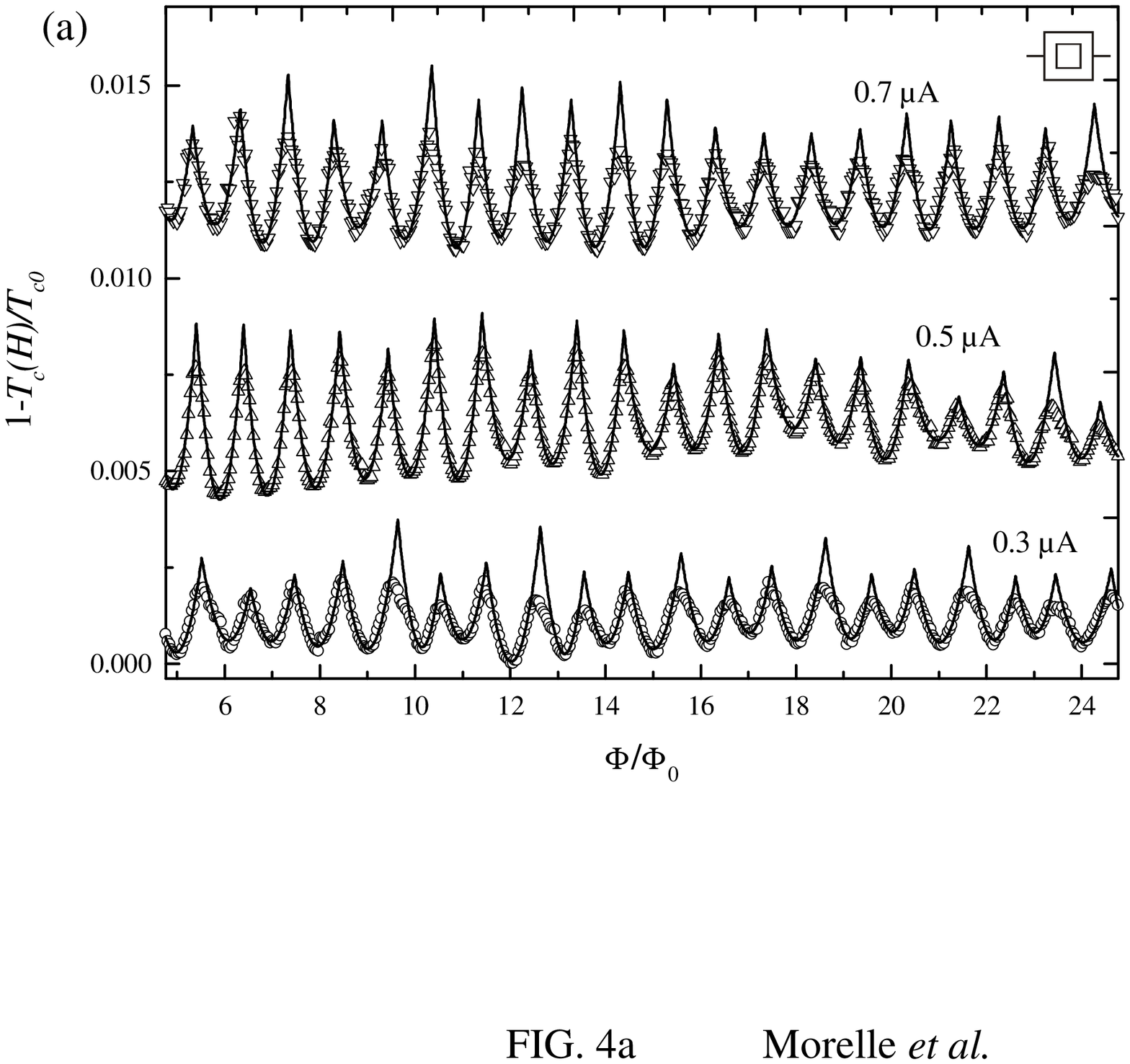}
\centering
\includegraphics*[width=8cm,clip=]{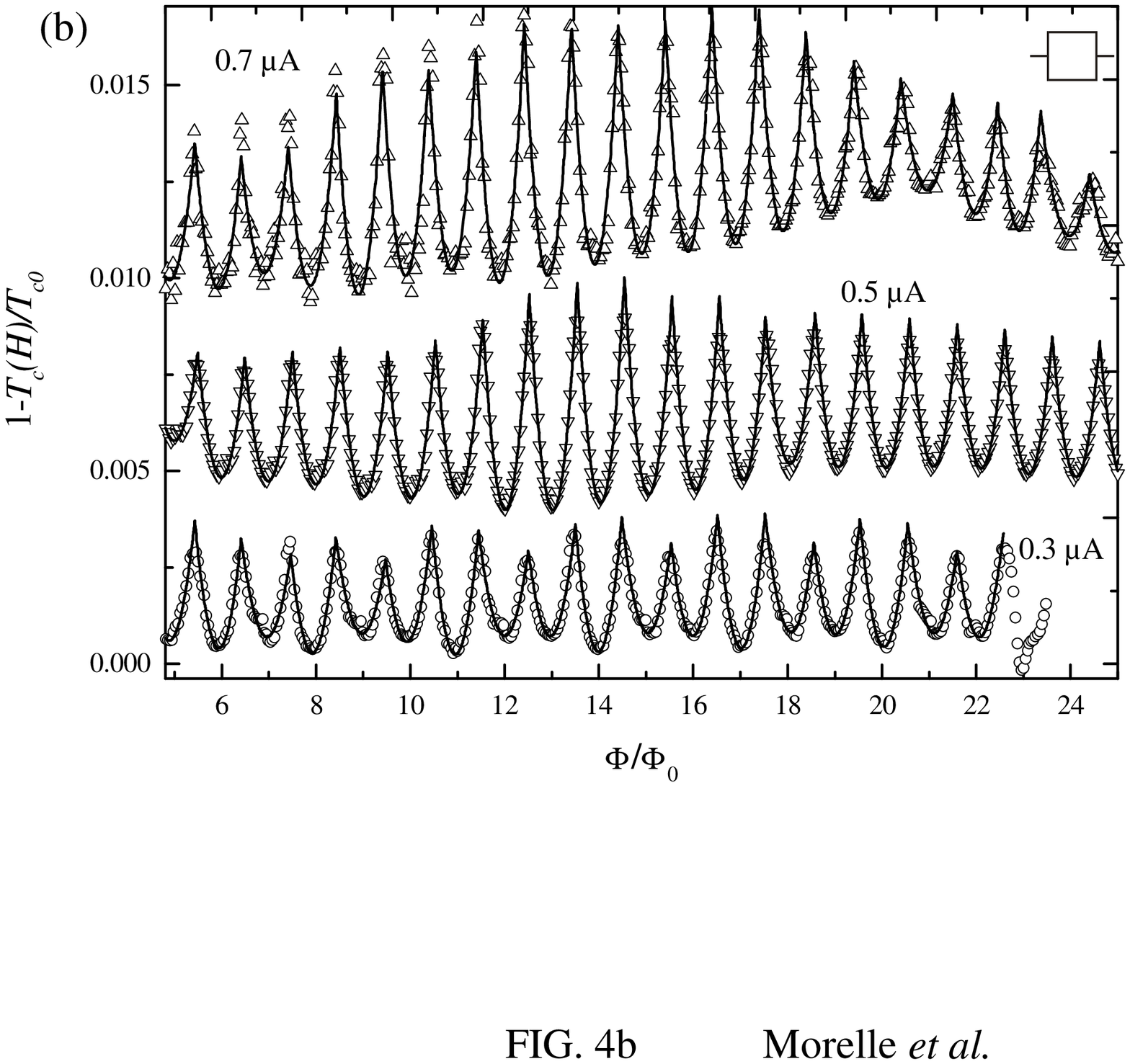}

\caption{Measured phase boundary $T_c(\Phi)$ after subtraction of
the parabolic background caused by the finite wire width of the
loops. The phase boundary is plotted in normalized units of the
flux $\Phi=S_o\mu_0 H$ threading the outer loop.  Within each
oscillation period, a parabolic function is fitted through the
data points in between the transition to a different fluxoid
quantum number $N_o$. The results for three different ac transport
currents $I_t$ ($\bigcirc$: 0.3 $\mu$A, $\bigtriangleup$:
0.5$\mu$A, $\bigtriangledown$: 0.7 $\mu$A) are presented (a) for
sample A and (b) for sample R.} \label{Fig:4}
\end{figure}

The phase boundaries of the double loop sample A and for the
reference sample R are also measured for different transport
currents $I_t$. The results for three different ac transport
currents $I_t$ ($\bigcirc$: 0.3 $\mu$A, $\bigtriangleup$: 0.5
$\mu$A, $\bigtriangledown$: 0.7 $\mu$A rms) are shown in Fig.
\ref{Fig:4}a and Fig. \ref{Fig:4}b for sample A and sample R,
respectively. From these experimental data, clear differences
between sample A and sample R are seen. First of all, the
amplitude of the $T_c(\Phi)$ oscillations is stronger in sample R,
secondly the phase boundary of the single loop (Fig. \ref{Fig:4}b)
matches very well with the fitted parabolic curves (solid curves),
while the oscillations of sample A (Fig. \ref{Fig:4}a) are not
parabolic at all, since the cusps in $T_c(\Phi)$ are always
rounded.

\begin{figure}[htb]
\centering
\includegraphics*[width=8cm,clip=]{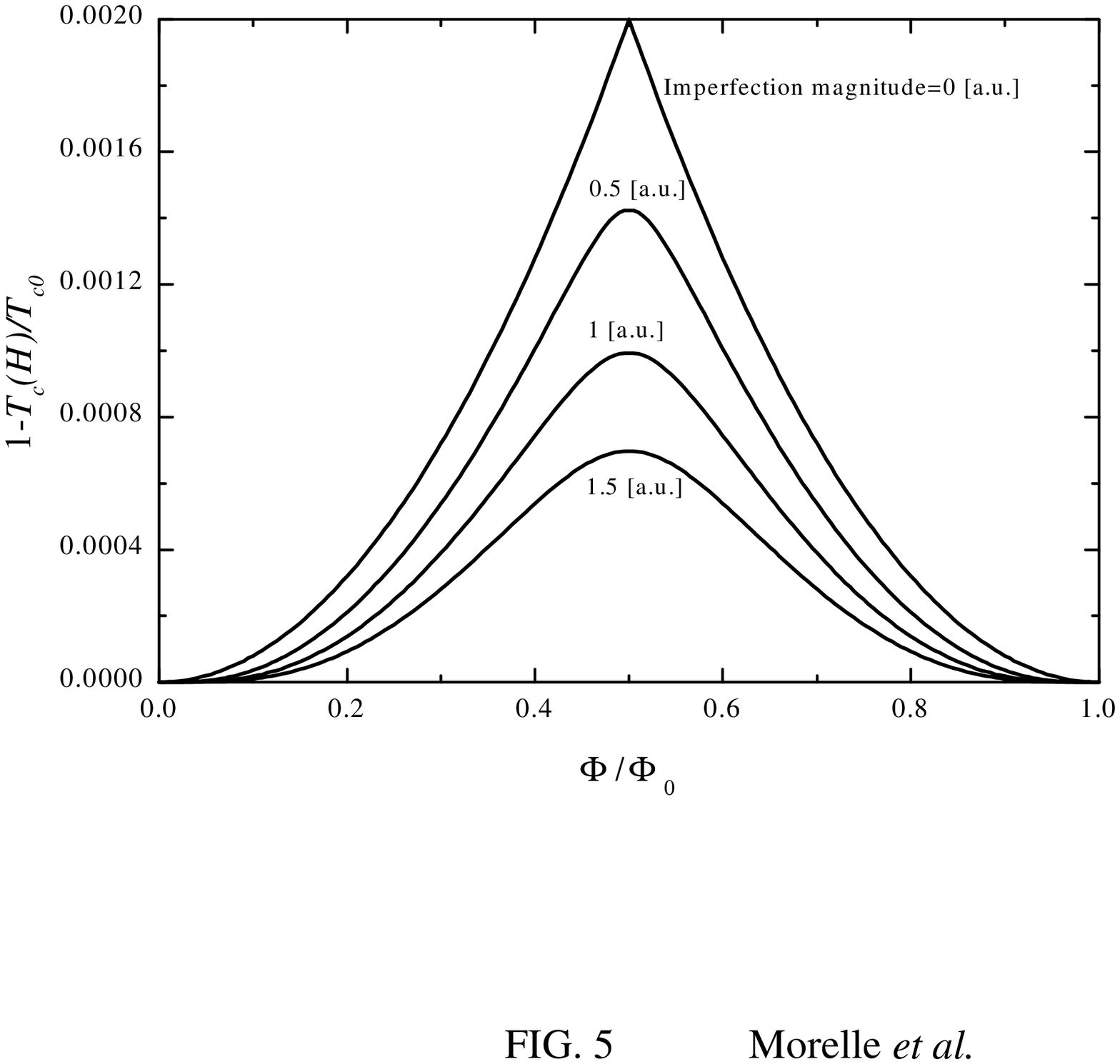}
\caption{The calculated phase boundary is shown for one period
using the micronet approach for a loop containing one scattering
imperfection\cite{Fomin}. The different curves correspond to
different magnitudes of the scattering imperfection: from zero
magnitude for the upper curve to a large magnitude for the lowest
curve.} \label{Fig:5}
\end{figure}

The rounding of the cusps in $T_c(\Phi)$, observed for sample A,
are not reproduced in samples B and C (see Fig. \ref{Fig:3}).
These rounded cusps cannot be attributed with all certainty to the
magnetic interactions with the inner loop, but may be also related
to the presence of scattering imperfections in the outer loop of
sample A, as has been shown theoretically in ref\cite{Fomin}. A
possible source of such imperfections could be the variation of
the strip width along the loops written by e-beam lithography.
Using the micronet approach\cite{Fink,Alexander}, the GL equation
can also be solved when imperfections are present in a loop. A
single period of the phase boundary calculated from this micronet
approach\cite{Fomin} is presented in Fig. \ref{Fig:5} for
different magnitudes of imperfections. Notice that the stronger
the imperfection the more the oscillation amplitude is damped.

\begin{figure}[htb]
\centering
\includegraphics*[width=8cm,clip=]{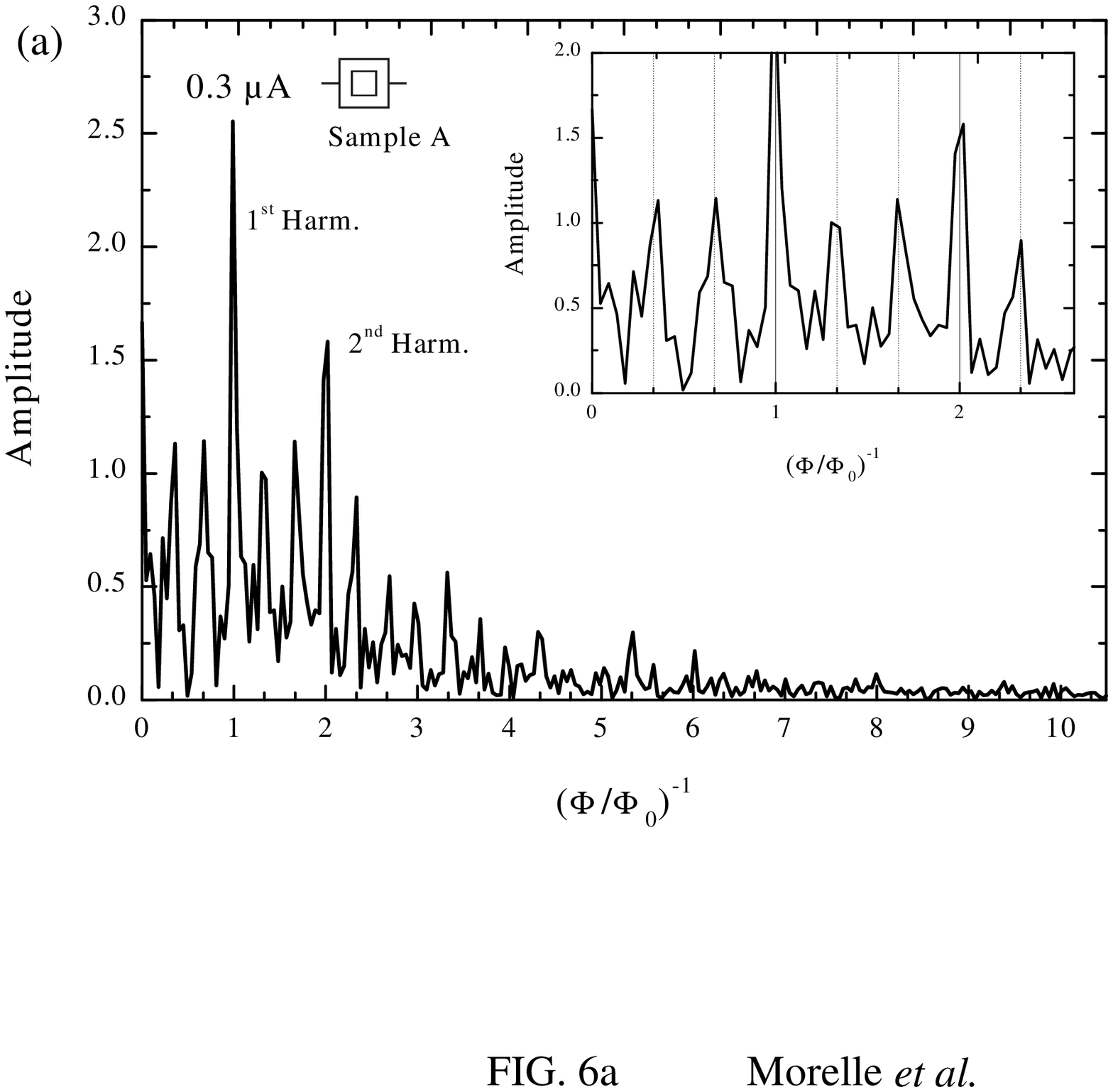}
\centering
\includegraphics*[width=8cm,clip=]{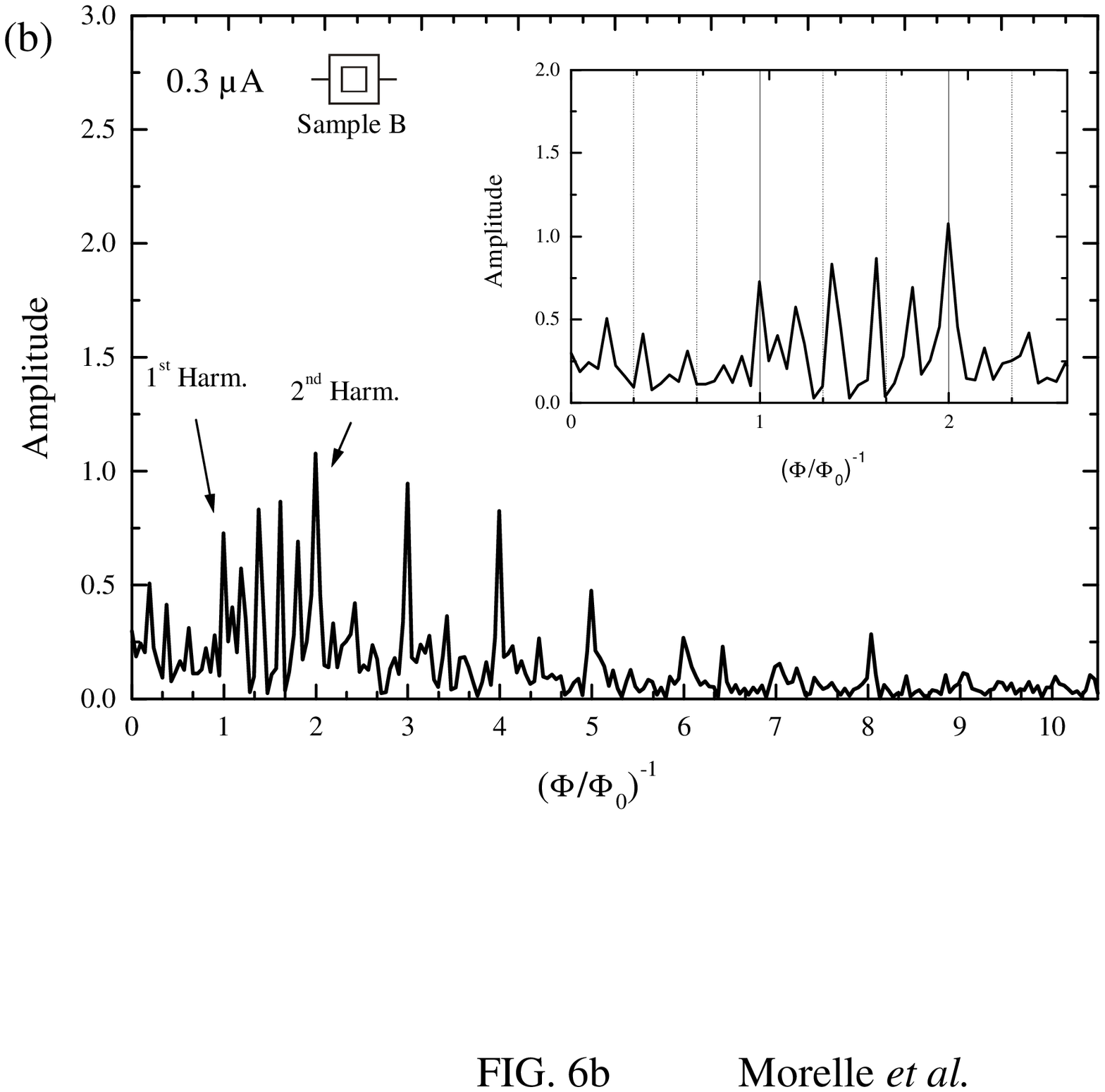}
\centering
\includegraphics*[width=8cm,clip=]{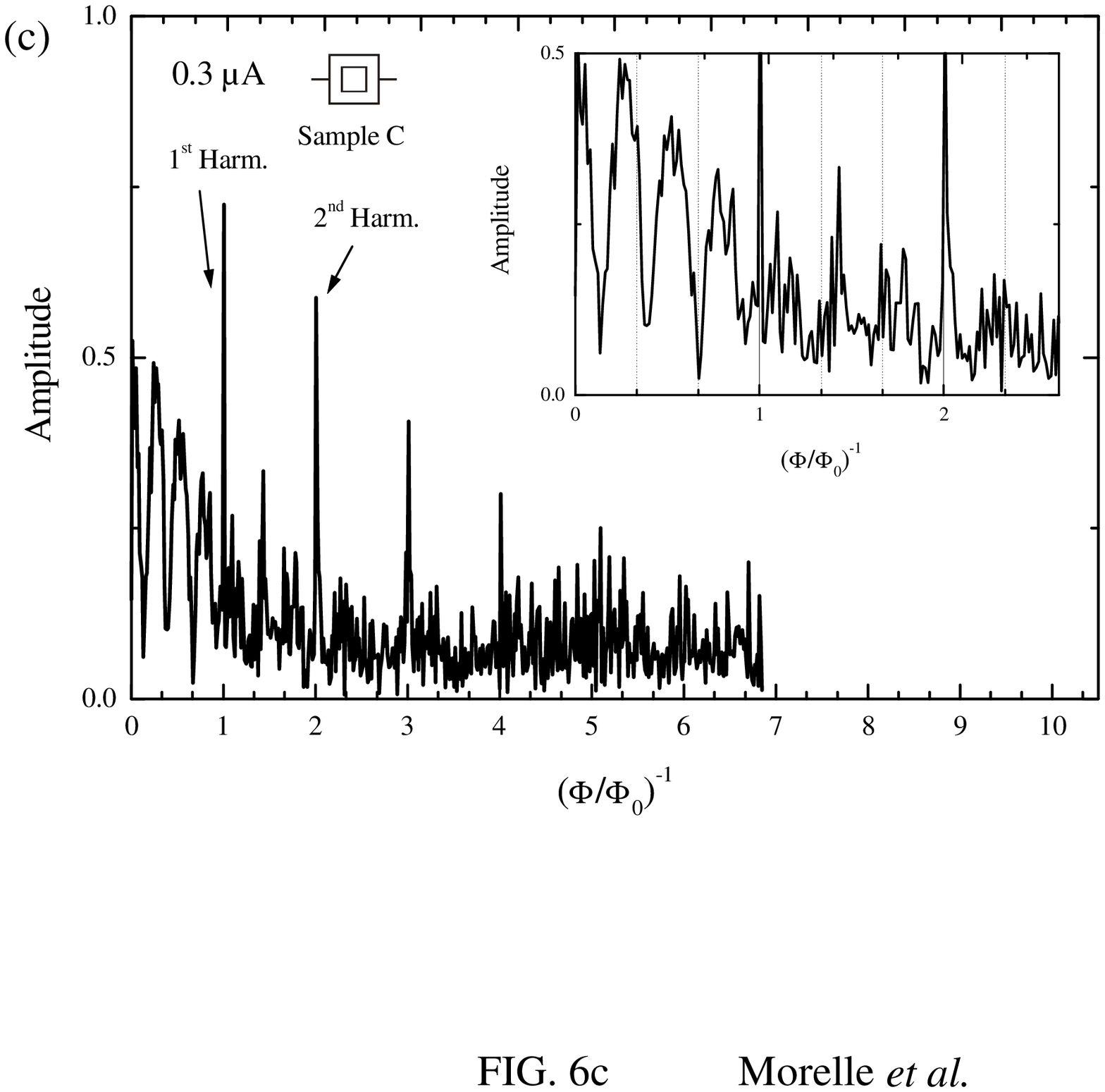}
\centering
\includegraphics*[width=8cm,clip=]{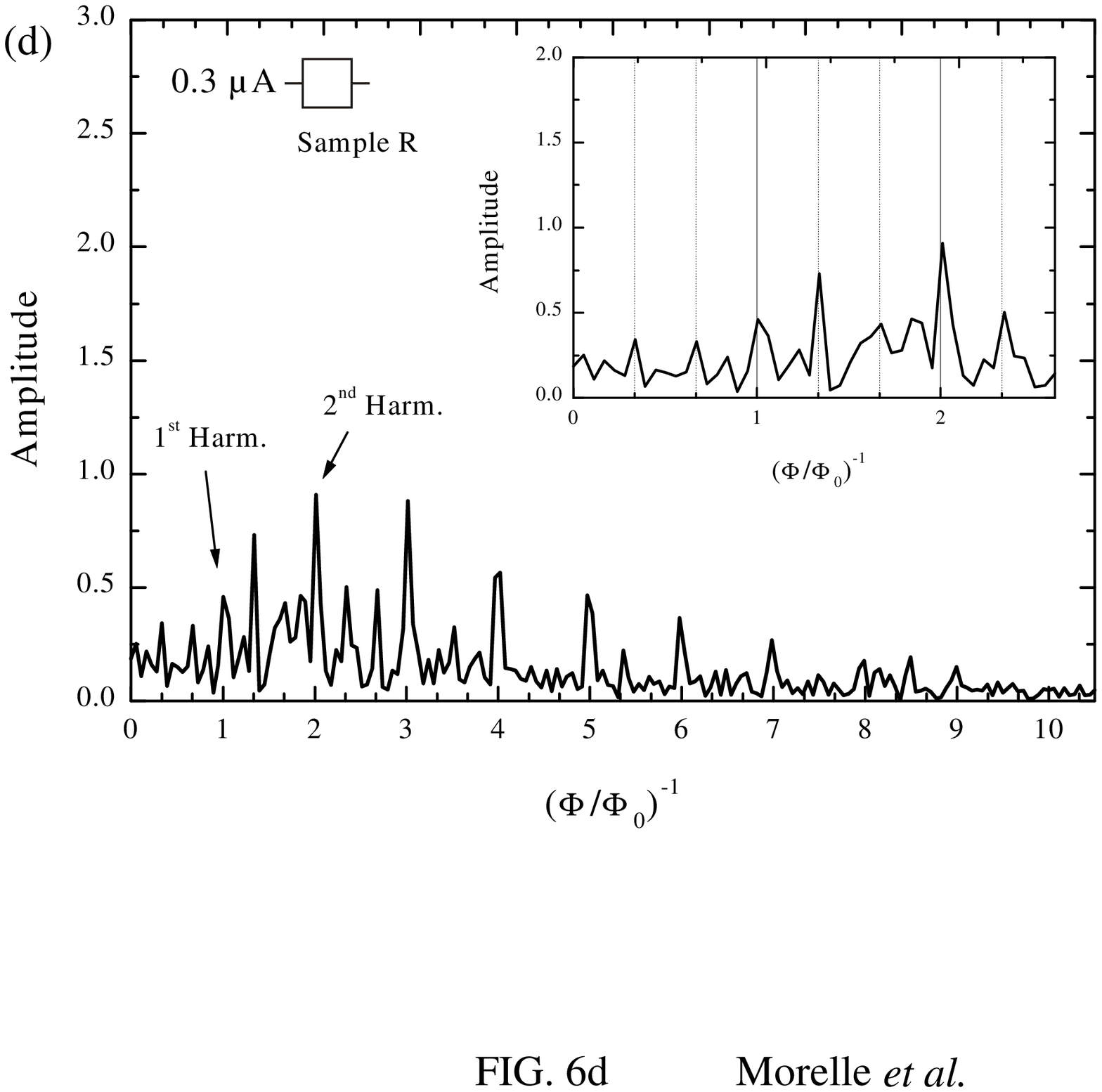}
\caption{Fourier transform of the phase boundary after subtraction
of the fitted parabolas within each oscillation period, for a
transport current $I_t$= 0.3 $\mu$A for (a) sample A, (b) sample
B, (c) sample C and (d) sample R. The inset shows a zoom of the
plot for the low frequency region.} \label{Fig:6}
\end{figure}

\begin{figure}[htb]
\centering
\includegraphics*[width=8cm,clip=]{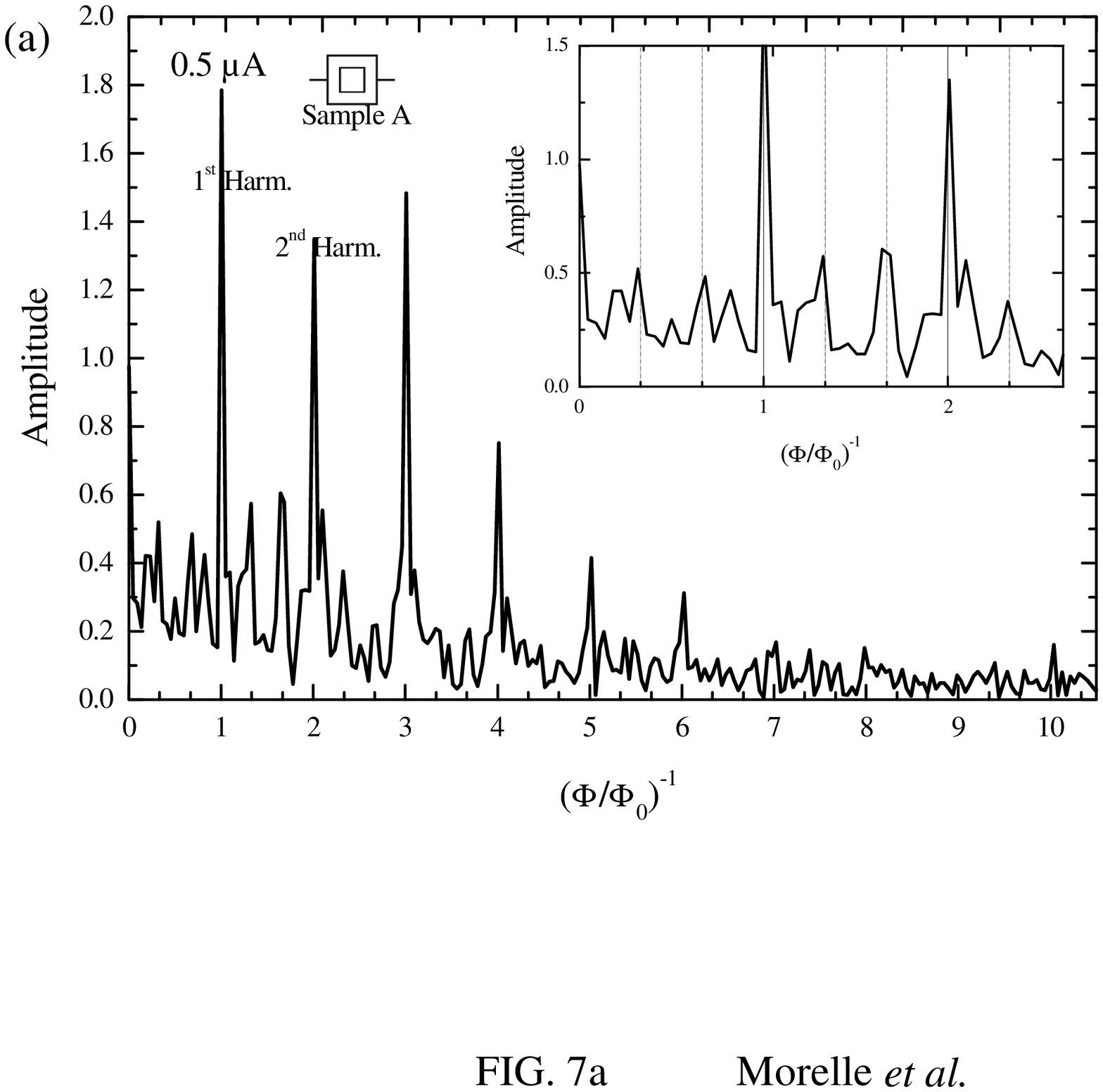}
\centering
\includegraphics*[width=8cm,clip=]{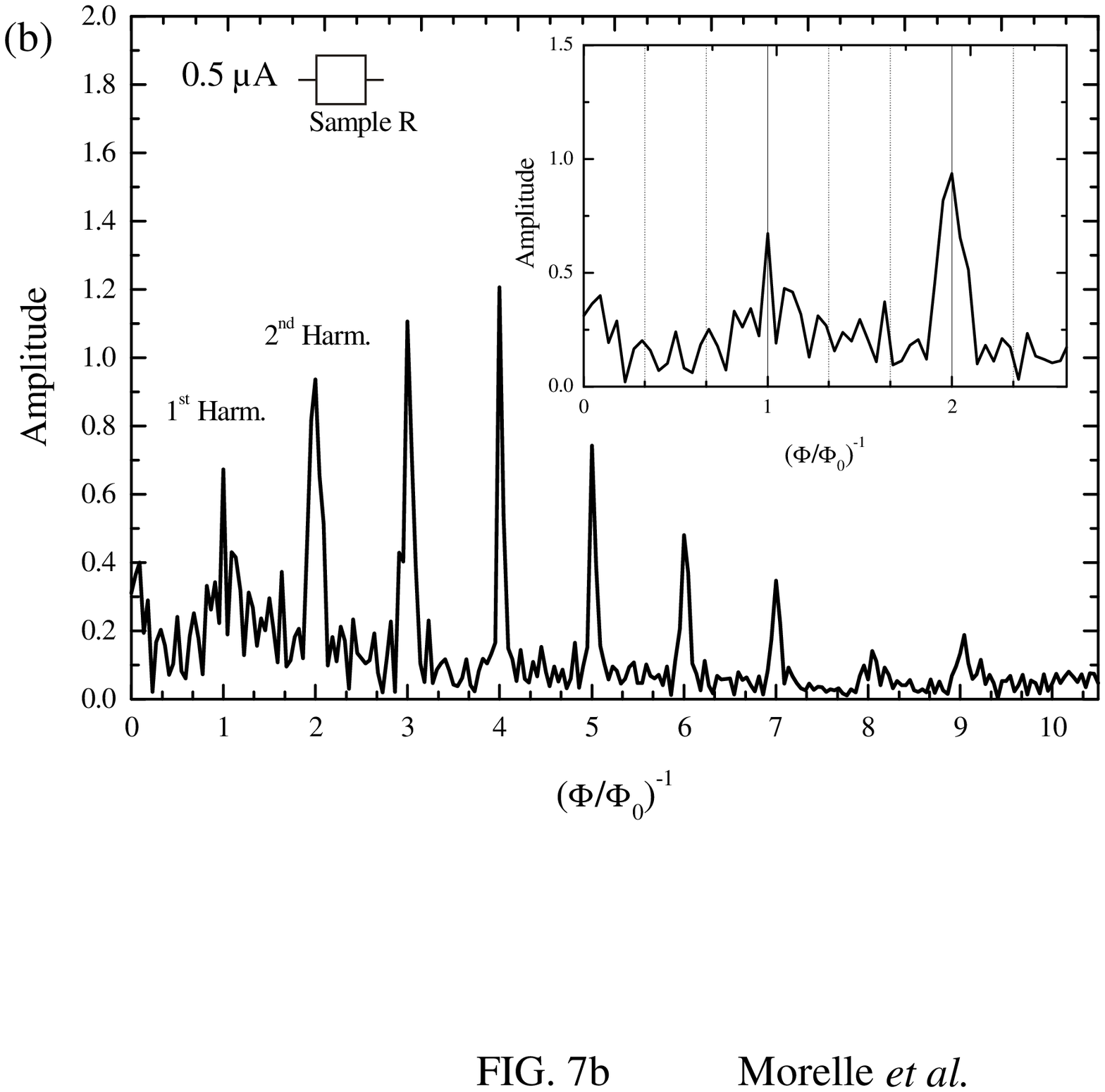}
\caption{Fourier transform of the phase boundary after subtraction
of the fitted parabolas within each oscillation period, for a
transport current $I_t$= 0.5 $\mu$A for (a) sample A and (b)
sample R. The inset shows a zoom of the plot for the low frequency
region.} \label{Fig:7}
\end{figure}

It is clear from our experiments and from\cite{Zhang} that the
influence of the inner loop on the measured phase boundary is
quite small. To detect traces of the periodicity coming from the
$T_c(\Phi)$ oscillations of the inner loop, a fast Fourier
transform analysis of the phase boundaries is carried out using
$2^{10}$ equally spaced points in an interval between 5 and 24
$\Phi/\Phi_0$ for sample A, B and R and between -36 and 49
$\Phi/\Phi_0$ for sample C, but with approximately the same number
of data points. In order to remove most of the contribution
arising from the fluxoid quantization in the outer loop, the
parabolas fitted in Fig. \ref{Fig:3}, \ref{Fig:4}a and
\ref{Fig:4}b (solid line) are subtracted from the experimental
data, prior to taking the Fourier transform. The resulting spectra
are presented in Fig. \ref{Fig:6}a for sample A, \ref{Fig:6}b for
sample B, \ref{Fig:6}c for sample C and \ref{Fig:6}d for the
single loop (sample R), for an external transport current of
$I_t$=0.3 $\mu$A. The Fourier spectra of sample A and sample R for
$I_t$=0.5 $\mu$A are shown in Fig. \ref{Fig:7}a and \ref{Fig:7}b,
respectively.

In Fig. \ref{Fig:6}a, \ref{Fig:6}c and \ref{Fig:7}a, a peak is
clearly seen at $(\Phi/\Phi_0)^{-1}=1$ corresponding to the
frequency of the LP oscillations for the outer loop; the second to
the sixth harmonics of this base frequency are also seen. In the
case of the sample R (Fig. \ref{Fig:6}d and \ref{Fig:7}b) and of
sample B (Fig. \ref{Fig:6}b), the first harmonic is strongly
reduced. This reduction is quite reasonable because of the
subtraction procedure applied before the Fourier transformation.
The lower scale for the Fourier spectrum for sample C (Fig.
\ref{Fig:6}c) is probably due to a smaller amplitude of the
oscillations in the phase boundary. For the double loop (sample A,
B and C), supplementary peaks are clearly distinguished in between
the harmonics (see inset in Fig. \ref{Fig:6}a, \ref{Fig:6}b,
\ref{Fig:6}c and \ref{Fig:7}a), which are considerably weaker in
the reference sample R. These peaks can be due to the coupling
with the supercurrent in the inner loop. Comparing the insets of
Fig. \ref{Fig:6}b and Fig. \ref{Fig:6}d, we note that the
supplementary peaks are substantially sharper and higher for
sample B than for sample R.

The period of the LP oscillations for the outer loop is $\mu_0
\Delta H_o=\Phi_0/S_o$=0.525 mT and $\mu_0 \Delta H_o$=0.550 mT
for sample A, B and R and for sample C, respectively,
corresponding to the first harmonic peak at
$(\Phi/\Phi_0)^{-1}=1$. The periodicity of the $T_c(\Phi)$
oscillations for the inner loop (samples A and B) is $\mu_0\Delta
H_i=\Phi_0/S_i$=2.6 mT, which is approximately 5 times larger than
$\mu_0\Delta H_o$, since the surface of the inner loop is
approximately 5 times smaller than the one of the outer loop. This
periodicity is in good agreement with the measurement on sample B
(Fig. \ref{Fig:6}b), where four peaks are indeed observed between
the first and the second harmonic, thus indicating a five times
smaller frequency compared to the frequency of the $T_c(\Phi)$
oscillations of the outer loop. The $T_c(\Phi)$ measurements on
sample A show two pronounced peaks between each harmonic (Fig.
\ref{Fig:6}a and \ref{Fig:7}a). This suggests a period for the
inner loop $\mu_0\Delta H_i\approx3\mu_0\Delta H_o$ which is in
disagreement with the dimensions of sample A. For sample C, which
has a slightly different size, the periodicity of the inner loop
oscillations in $T_c(\Phi)$ is $\mu_0\Delta H_i=\Phi_0/S_i$=2.1
mT$\approx 4 \mu_0\Delta H_o$. This periodicity is in agreement
with the measurements where three peaks are clearly observed
before the first harmonic (see Fig. \ref{Fig:6}c).

The reason of the disagreement between the periodicity of the
oscillations and the peaks in the Fourier spectrum of sample A can
be a different effective surface. To calculate the periodicity of
the oscillations of the inner and outer loop, the average size of
the loops (through the middle of the wires) has been used. It is
possible that we have to take a slightly smaller or larger
effective surface for sample A. The dimensions of samples A and B
may also be not exactly the same. But the effective surface has to
be taken larger than the largest dimensions of the inner loop to
obtain a value of $\mu_0\Delta H_i\approx3\mu_0\Delta H_o$. We
therefore think that the smaller peaks might be hidden, and not
clearly seen however in the Fourier spectrum. In that case, a
value for the periodicity of 6 times the periodicity of the inner
loop is more realistic for sample A. Coming back to Fig.
\ref{Fig:2}, we can see that the interval (between 5 and 24
$\Phi/\Phi_0$), where a Fourier transform was performed for sample
A, B and R, corresponds only to 3 or 4 periods $\mu_0\Delta H_i$.
The resolution of the Fourier spectrum in the low frequency regime
will therefore be low, thus presenting an additional difficulty in
interpreting the intermediate peaks in the Fourier spectra. The
phase boundary $T_c(\Phi)$ of Sample C is measured over a broader
interval (between -36 to 49 $\Phi/\Phi_0$). This interval
corresponds to more than 20 periods $\mu_0\Delta H_i$, what
results in a better resolution of the Fourier spectrum in the low
frequency regime.

In a first approximation, with the currents $I_i$ and $I_o$
independent from each other, the mutual inductance $M$ can be
evaluated from the amplitude of the additional peaks in the
Fourier spectrum. The additional energy due to the mutual
inductance in Eq.~\ref{Eq:Freeenergy} ($MI_iI_o$) has to be of the
same order of magnitude as the amplitude of the oscillations
$k_B\Delta T_{coupling}$ due to coupling in the phase boundary of
the outer loop. The value for the mutual inductance can then be
evaluated with the formula
\begin{equation}
M \approx \frac{k_B\Delta T_{coupling}}{\langle I_iI_o\rangle},
\label{Eq:compare}
\end{equation}
with $k_B$ the Boltzmann constant. This gives $M \approx$ 17 pH,
13 pH and 14 pH for sample A, B and C, respectively. The average
of the supercurrents $I_i$ and $I_0$ is calculated from
ref\cite{Davidovic,Zhang}, for $I_t$=0.3~$\mu$A and with the
criterion for $T_{co}$ at 90\% of $R_n$, what correspond to a
typical temperature interval of 3 mK between this criterion and
the 50\% criterion used for the measurements of the phase
boundary. It is quite clear that the supercurrents, and thus also
the mutual inductance are strongly dependent of this chosen
criterion. The values calculated from Eq.~\ref{Eq:compare} are
comparable to the calculated mutual inductances from table
\ref{table:param}. Due to the strong temperature dependent
currents, only the order of magnitude of the mutual inductance can
be evaluated from our measurements.

For high ($I_t$=0.9 $\mu$A for sample A and 1.0 $\mu$A for sample
C) transport current, the peaks between the harmonics completely
disappear (not shown). This vanishing may be due to increasing
shift of $T_c(\Phi)$ with the applied transport current. The
highest current corresponds to the case of the upper curve
(dashed-dotted line) in Fig. \ref{Fig:2}. Hence, for all flux
values, the inner loop is in the superconducting state at
$T_c(\Phi)$ of the outer loop, and the phase boundaries of the two
loops are not intersecting each other. There will be no sharp
interruption of the supercurrent of the inner loop going from the
superconducting to the normal state. Therefore, the shape of the
phase boundary of the outer loop can be less sensitive to the
presence of the inner loop. On the other hand, once the inner loop
is deep in the superconducting state, a higher supercurrent in the
inner loop would be present, and this higher supercurrent would
require a higher coupling. If the transport current $I_t$ is high
enough, the inner loop will be always superconducting at
$T_c(\Phi)$ of the outer loop. A discontinuity in the measured
phase boundary of the outer loop is expected when the inner loop
changes from a fluxoid quantum number $N_i$ to $N_i\pm1$. However,
we could not see a discontinuity in the measured phase boundaries,
corresponding to a sign reversal of the supercurrent in the inner
loop.

\section{Conclusion}

We have measured the normal/superconducting phase boundary of a
superconducting system consisting of two concentric mesoscopic
loops, to study magnetic interactions between the two loops.  The
modification of the $T_c(\Phi)$ oscillations of the outer loop is
seen in the Fourier spectrum of the $T_c(\Phi)$ line due to the
coupling between the outer and the inner loops. To interpret these
observations, we have used two different models. The first model
assumes the presence of scattering imperfections in the outer
loop. This model cannot explain the observed evolution of the
Fourier spectrum with the current, although it might be applicable
for a fixed weak current. The second model explains the extra
peaks in the Fourier spectrum by the magnetic coupling of the two
loops. The systematic shift of the $T_c(\Phi)$ phase boundary of
the outer loop with the applied current $I_t$ induces a well
defined evolution of the Fourier Spectrum which was indeed found
in our experiments. This evolution of the extra peaks in the
Fourier spectrum with the applied current gives an experimental
evidence for the presence of the magnetic interaction between the
two superconducting loops.

Future magnetic measurements on huge arrays of magnetically
coupled loops, deeper in the superconducting state, could be
helpful to reveal an enhanced magnetic coupling of both loops at
lower temperatures. A inner loop made from a different
superconductor with a higher critical temperature would certainly
increase the magnetic coupling between the two loops. An enhanced
critical field is expected in this case for the lower $T_c$ loop,
at the expense of sharing the fluxoid quantization "burden" with a
loop where superconductivity is stronger.

\section{Acknowledgments}
The authors wish to thank M. Cannaerts, E. Seynaeve and K. Temst
for the AFM, SEM and X-ray measurements, and H.J. Fink and L. Van
Look for useful discussions. This work has been supported by the
Belgian IUAP, the FWO and GOA-programs, and by the ESF Programme
VORTEX.

\end{document}